\documentstyle[11pt]{article}

\newcommand{\be}{\begin{equation}}
\newcommand{\ee}{\end{equation}}
\newcommand{\ea}{\end{eqnarray}}
\newcommand{\ba}{\begin{eqnarray}}
\def\ni{\noindent}

\begin{document}

\begin{titlepage}
\vspace{1cm}

\begin{centering}

{\Large \bf  Noncommutative Particles in Curved Spaces }

\vspace{.5cm}
\vspace{1cm}

{\large E. M. C. Abreu$^{a,b}$, \large R. Amorim$^{c}$} and {\large W. Guzm\'an Ram\'{\i}rez$^{b}$}

\vspace{0.5cm}

$^a$Grupo de F\' isica Te\'orica e Matem\'atica F\' isica,
Departamento de F\'{\i}sica,\\
 Universidade Federal Rural do Rio de Janeiro,\\
BR 465-07, 23890-971, Serop\'edica, Rio de Janeiro, Brazil\\[1.5ex]

$^b$ Centro Brasileiro de Pesquisas F\' isicas (CBPF),
Rua Dr. Xavier Sigaud 150, Urca, \\
CEP  22290-180, Rio de Janeiro, Brazil \\[1.5ex]

$^c$Instituto de F\'{\i}sica, Universidade Federal
do Rio de Janeiro,\\
Caixa Postal 68528, 21945-970,  Rio de Janeiro, Brazil\\[1.5ex]

\today

\vspace{1cm}

\begin{abstract}
\ni We present a formulation in a curved background of noncommutative mechanics, where the object of noncommutativity $\theta^{\mu\nu}$ is considered as an independent quantity having a canonical conjugate momentum.  We introduced a noncommutative first-order action in $D=10$ curved spacetime and the covariant equations of motions were computed.
This model,  invariant under diffeomorphism, generalizes recent  relativistic results. 
\end{abstract}

\end{centering}

\vspace{1cm}

\noindent Keywords: noncommutativity, extended DFR space, curved spaces \\

\noindent PACS: 04.40.-b; 11.10.Nx; 11.15.-q

\vfill

\noindent{\tt evertonabreu@ufrrj.br\\  amorim@if.ufrj.br\\ wguzman@cbpf.br}

\end{titlepage}

\pagebreak

\section{Introduction}
\renewcommand{\theequation}{1.\arabic{equation}}
\setcounter{equation}{0}

There are strong theoretical evidences that at Planck scale, the dominant geometrical scenario is the noncommutative (NC) one.  The fact that noncommutativity manifests itself as a consequence of string theory embedded into a magnetic field background \cite{SW} rekindled the interest in NC geometry \cite{sigma} which is considered as a fundamental element of quantum gravity  (for reviews see \cite{szabo,nekrasov}).

The main geometrical feature considered by General Relativity and Quantum Field Theory is that both assume that spacetime is a continuum.  Based on this assumption, a pseudo Riemannian manifold furnish the basis for the geometrical description of the general theory of relativity \cite{vw}.   In this way, quantum fields and their interactions are local operators that are functions of continuous spacetime coordinates.  However, to have an unified theory, it seems that such a continuous spacetime is not the best choice.  The paths described above pinpoint to a discrete spacetime structure, at the Planck scale, with NC coordinates.  The continuum spacetime is the limiting case.  This geometrical issue can also be fathomed since the unification of General Relativity (GR) and Quantum Mechanics (QM) implies the existence of a fundamental length \cite{cgh}.  The models that incorporate the notion of a fundamental length form the class of gauge theories formulated on NC space, which has a fundamental length defined in Planck scale, $\lambda_{Pl}$.

These spontaneous manifestations of noncommutativity can lead, for example, to the fact that standard four-dimensional spacetimes may become NC, namely, that the position four-vector operator ${\mathbf x}^\mu$ obeys the following rule
\be \label{a1}
[\,{\mathbf x}^\mu\,,\,{\mathbf x}^\nu\,]\,=\,i\,\theta^{\mu\nu}\,\,,
\ee

\ni where $\theta^{\mu\nu}$ is a real, antisymmetric and constant matrix.  It can be demonstrated that, in a certain limit, a gauge theory on noncommutative spaces is tantamount to string theory.    

In seminal works \cite{Snyder,yang}, it was introduced a five dimensional spacetime with SO(4,1) as a symmetry group, with generators ${\mathbf M}^{AB}$, satisfying the Lorentz algebra, where $A,B=0,1,2,3,4$ and using natural units, i.e., $\hbar=c=1$. Moreover, the relation between coordinates and  generators of the $SO(4,1)$ algebra was written as,
$$\,\,
{\mathbf x}^{\mu}=a\,{\mathbf M}^{4\mu}\,\,
$$
(where $\mu,\nu=0,1,2,3$ and the parameter $a$ has dimension of length), promoting in this way the spacetime coordinates to Hermitian operators. The mentioned relation introduced the commutator,
\begin{equation}
\label{2}
[{\mathbf x}^\mu,{\mathbf x}^\nu] = i a^2{\mathbf M}^{\mu\nu}
\end{equation}

\noindent and the identities,
\be \label{2.1}
\,[{\mathbf M}^{\mu\nu},{\mathbf x}^\lambda]= i({\mathbf x}^{\mu}\eta^{\nu\lambda}-{\mathbf x}^{\nu}\eta^{\mu\lambda})\,
\ee
and
\be \label{2.2}
\,[{\mathbf M}^{\mu\nu},{\mathbf M}^{\alpha\beta}]= i({\mathbf M}^{\mu\beta}\eta^{\nu\alpha}-{\mathbf M}^{\mu\alpha}\eta^{\nu\beta}+{\mathbf M}^{\nu\alpha}\eta^{\mu\beta}-{\mathbf M}^{\nu\beta}\eta^{\mu\alpha})\,,
\ee
which agree with four dimensional Lorentz invariance. 

Some years back, in \cite{DFR}, the authors essentially assumes 
(\ref{a1}) as well as the vanishing of the triple commutator among the coordinate operators.  The Doplicher-Fredenhagen-Roberts (DFR) algebra is based on  principles imported from GR and QM. In addition to (\ref{a1}) it also assumes that 
\begin{equation}
\label{6}
[{\mathbf x}^\mu,{\mathbf \theta}^{\alpha\beta}] = 0\,\,.
\end{equation}

With this formalism, DFR demonstrated that after the combination of QM with classical gravitation theory, the ordinary spacetime loses all operational meaning at short distances.

An important point in DFR algebra is that the Weyl representation of NC operators obeying 
(\ref{a1}) and (\ref{6}) keeps the usual form of the Moyal product, 
\begin{equation} \label{a11.22}
\varphi(x)\,\star\,\psi(x)\,=\,exp\Big[\frac{i}{2}\,\theta^{\mu\nu}\,\frac{\partial}{\partial\,x^{\mu}}\frac{\partial}{\partial\,y^{\nu}} \Big]\,\varphi(x)\,\psi(y)\mid_{x=y}
\end{equation}
and consequently the form of the usual NCFT's, although the fields have to be considered as dependent not only on ${\mathbf x}^\mu$ but also on ${\mathbf \theta}^{\alpha\beta}$.
The argument is that  very accurate measurements of spacetime localization could transfer to test particles, sufficient energies to create a gravitational field that in principle could trap photons. This possibility is related to spacetime uncertainty relations that can be derived from (\ref{a1}) and (\ref{6}) as well as from the quantum conditions 
\begin{eqnarray}
\label{04000}
&&{\mathbf \theta}_{\mu\nu}{\mathbf \theta}^{\mu\nu} =0\nonumber\\
&&({1\over4}\;{}^*{\mathbf \theta}^{\mu\nu}{\mathbf \theta}_{\mu\nu})^2=\lambda_{Pl}^8
\end{eqnarray}

\noindent where $^*{\mathbf \theta}_{\mu\nu}={1\over2}\epsilon_{\mu\nu\rho\sigma}{\mathbf \theta}^{\rho\sigma}$.

In \cite{Amorim1}, one of us promoted a DFR algebra extension to a non-relativistic QM in the trivial way, but keeping consistency. 
The objects of noncommutativity were considered as  true operators and their conjugate momenta were introduced. This permits to display a complete and consistent algebra among the Hilbert space operators and to construct generalized angular momentum operators, obeying the $SO(D)$ algebra, and in a dynamical way, acting properly in all the sectors of the Hilbert space.  If this is not accomplished, some fundamental objects usually employed in the literature, as the shifted coordinate operator (see (\ref{16000}) below), fail to properly transform under rotations. Symmetry is implemented not in a mere algebraic way, where the transformations are based on structure indices of the variables.  But it comes dynamically from the consistent action of an operator, as discussed in \cite{Iorio}.  This new NC space has ten dimensions and it has been called since then, the extended DFR space.  Next section we will review details of this new NC space.

Recently \cite{Amorim6}, the formalism presented in \cite{Amorim4} was generalized to the relativistic case.  A noncommutative relativistic classical theory was constructed which, under quantization, furnishes the theories presented 
in \cite{Amorim2,Amorim5}.  The results obtained there, are invariant under the action of the Lorentz group SO(1,D) as well as under some generalization of the Poincar\'e group \cite{Amorim2,Amorim3}.  The relation between the formalism in \cite{Amorim6} is related to the one in \cite{Deriglazov}, after the elimination of some auxiliary variables.

Analyzing the theory of quantum fields on curved spacetimes one realize that gravity is considered as classical background and discuss the features of these quantum fields propagating on this background.  The structure of spacetime is described by a manifold ${\cal M}$ with metric $g_{\mu\nu}$, i.e., a $({\cal M},g)$ commutative curved spacetime.  In $({\cal M},g)$ there is a wide variety of interesting phenomena, such as particle creation near a black hole with the Schwarzschild radius much greater than the Planck length.  We will give more details about $({\cal M},g)$ in the third section.

The main motivation to study NC systems in curved spacetimes is to make contact to physics such as cosmic microwave background or Hawking radiation \cite{su}.  However, the majority of these studies make use of the Moyal-Weyl (the star product) or $\kappa$-deformed Minkowski spacetime.  Other formalisms using classical methods has been applied in order to introduce noncommutativity in gravity systems, as shifted coordinates and symplectic deformations \cite{po,gss}.

In this work we perform an extension of the system analyzed in \cite{Amorim6} embedding it in a curved 
space \cite{Amorim}.  We used the vielbein formalism and a set of auxiliary variables, the so-called Einstein-Kramer \cite{ek,Amorim} variables in order to write the action in a generalized gravitational background.  As we said, the extended DFR space is a larger one with 10 dimensions, 4 for the spacetime and 6 for the $\theta$-space.  The analysis concerning dimension reduction formalisms like Kaluza-Klein or holography are out of the scope of this work, although are targets for future investigations.  The equations of motion indicating the dynamics of this system in curved space were obtained.

The paper is organized such that in section 2 we review the quantum mechanics in the extended DFR space.  In section 3 we describe the main steps of curved spaces formalism and its variational principles were discussed in order to introduce the main ingredients that will be used throughout the paper.  In section 4 we present the mechanical system that will be embedded in a generalized gravitational background and its equations of motion will be physically discussed.  Comments and perspectives will be depicted in section 5, namely, in the conclusion section.

\section{Quantum Mechanics in the extended DFR noncommutative space}
\renewcommand{\theequation}{2.\arabic{equation}}
\setcounter{equation}{0}

In this section we will furnish the main ingredients of the extended DFR algebra.  The interested reader can find more details and an extensive list of noncommutativity reviews in \cite{sigma}.

The DFR algebra \cite{DFR} essentially assumes (\ref{a1}) as well as the vanishing of the triple commutator among the coordinate operators,
\be \label{triple}
[\,{\mathbf x}^{\mu}\,,\,[{\mathbf x}^{\nu}\,,\,{\mathbf x}^{\rho}\,]\,]\,=\,0\,\,.
\ee
It is easy to realize that this relation constitute a constraint in a NC spacetime.  Notice that the commutator inside the triple one is not a $c$-number.

The basic DFR algebra rely on  principles imported from GR and QM. In addition to (\ref{a1}) it also assumes that 
\begin{equation}
\label{6a}
[{\mathbf x}^\mu,{\mathbf \theta}^{\alpha\beta}] = 0\,\,,
\end{equation}

\ni and we consider that space has arbitrary $D\geq2$ dimensions.  As usual ${\mathbf x}^\mu$ and ${\mathbf p}_\nu$, where $i,j=1,2,...,D$,  and  $\mu,\nu=0,1,....,D$,  represent the position operator and its
conjugate momentum in Euclidean and Minkowski spaces respectively.   The NC variable ${\mathbf \theta}^{\mu\nu}$ represent the noncommutativity operator, but now ${\mathbf \pi}_{\mu\nu}$ is its conjugate momentum. In accordance with the discussion above, it follows the algebra
\begin{equation}
\label{7000}
[{\mathbf x}^\mu,{\mathbf p}_\nu] = i \delta^\mu_\nu\,\,,
\,\,\,\,\,\,\,\qquad \qquad
[{\mathbf \theta}^{\mu\nu},{\mathbf \pi}_{\alpha\beta}] = i \delta^{\mu\nu}_{\,\,\,\,\alpha\beta}
\end{equation} 

\noindent where $\delta^{\mu\nu}_{\,\,\,\,\alpha\beta}=\delta^{\mu}_{\alpha}\delta^{\nu}_{\beta}-\delta^{\mu}_{\beta}\delta^{\nu}_{\alpha}$. 
The relation (\ref{a1}) here in a space with $D$ dimensions, for example, can be written as
\begin{equation}
\label{9000}
[{\mathbf x}^i,{\mathbf x}^j] = i \,{\mathbf \theta}^{ij} \qquad \mbox{and} \qquad [{\mathbf p}_i,{\mathbf p}_j ] = 0
\end{equation}

\noindent and together with the triple commutator (\ref{triple}) condition of the standard spacetime, i.e., 
\begin{equation}
\label{10}
[{\mathbf x}^\mu,{\mathbf \theta}^{\nu\alpha}] = 0\,\,.
\end{equation}

\noindent This implies that 
\begin{equation}
\label{11}
[{\mathbf \theta}^{\mu\nu},{\mathbf \theta}^{\alpha\beta}] = 0\,\,,
\end{equation} 
and this completes the DFR algebra.

Recently, in order to obtain consistency it was introduced \cite{Amorim1}, as we talked above, the canonical conjugate momenta $\pi_{\mu\nu}$ such that,
\begin{equation}
\label{12}
[{\mathbf p}_\mu,{\mathbf \theta}^{\nu\alpha}] = 0\,\,,\,\,\,\,\,\,
[{\mathbf p}_\mu,{\mathbf \pi}_{\nu\alpha}] = 0\,\,.
\end{equation}

The Jacobi identity formed by the operators ${\mathbf x}^i$, ${\mathbf x}^j$ and ${\mathbf \pi}_{kl}$ leads to the nontrivial relation
\begin{equation}
\label{14}
[[{\mathbf x}^\mu,{\mathbf \pi}_{\alpha\beta}],{\mathbf x}^\nu]- [[{\mathbf x}^\nu,{\mathbf \pi}_{\alpha\beta}],{\mathbf x}^\mu]   =   - \delta^{\mu\nu}_{\,\,\,\,\alpha\beta}\,\,.
\end{equation}

\noindent The solution, unless trivial terms,  is given by
\begin{equation}
\label{15000}
[{\mathbf x}^\mu,{\mathbf \pi}_{\alpha\beta}]=-{i\over 2}\delta^{\mu\nu}_{\,\,\,\,\alpha\beta}{\mathbf p}_\nu \,\,.
\end{equation}

\noindent It is simple to verify that the whole set of commutation relations listed above is indeed consistent under all possible Jacobi identities. Expression (\ref{15000}) suggests the  shifted coordinate 
operator \cite{Chaichan,Gamboa,Kokado,Kijanka,Calmet}
\begin{equation}
\label{16000}
{\mathbf X}^\mu\equiv{\mathbf x}^\mu\,+\,{1\over 2}{\mathbf \theta}^{\mu\nu}{\mathbf p}_\nu\,\,,
\end{equation}

\noindent that commutes with ${\mathbf \pi}_{kl}$. Actually, (\ref{16000}) also commutes with ${\mathbf \theta}^{kl}$ and $ {\mathbf X}^j $, and satisfies a non trivial commutation relation with  ${\mathbf p}_i$  depending objects, which could be derived from
\begin{equation}
\label{17000}
[{\mathbf X}^\mu,{\mathbf p}_\nu]=i\delta^\mu_\nu
\end{equation}
\ni and
\begin{equation}
\label{i6}
[{\mathbf X}^\mu,{\mathbf X}^\nu]=0\,\,.
\end{equation}

To construct an extended DFR algebra in $(x,\theta)$ space, we can write 

$${\mathbf M}^{\mu\nu}\,=\,{\mathbf X}^\mu{\mathbf p}^\nu\,-\,{\mathbf X}^\nu{\mathbf p}^\mu\,-\,\theta^{\mu\sigma}\,\pi_{\sigma}^{\:\:\nu}\,+\,\theta^{\nu\sigma}\,\pi_{\sigma}^{\:\:\mu}\,\,,$$ where ${\mathbf M}^{\mu\nu}$ is the antisymmetric generator of the Lorentz-group.  To construct  $\pi_{\mu\nu}$ we have to obey equations
(\ref{7000}b) and (2.9), obviously.   From (\ref{7000}a) we can write the generators of translations as $$P_\mu = - i \partial_\mu\,\,.$$  With these ingredients it is easy to construct the commutation relations

\begin{eqnarray} \label{ABC}
\left[ {\mathbf P}_\mu , {\mathbf P}_\nu \right] &=& 0 \nonumber \\
\left[ {\mathbf M}_{\mu\nu},{\mathbf P}_{\rho} \right] &=& -\,i\,\big(\eta_{\mu\nu}\,{\mathbf P}_\rho\,-\,\eta_{\mu\rho}\,{\mathbf P}_\nu \big) \\
\left[ {\mathbf M}_{\mu\nu},{\mathbf M}_{\rho\sigma} \right] &=& -\,i\,(\,\eta_{\mu\rho}\,{\mathbf M}_{\nu\sigma}\,-\,\eta_{\mu\sigma}\,{\mathbf M}_{\nu\rho}\,-\,\eta_{\nu\rho}\,{\mathbf M}_{\mu\sigma}\,-\,\eta_{\nu\sigma}\,{\mathbf M}_{\mu\rho}\,)\;\;, \nonumber 
\end{eqnarray}
and we can say that ${\mathbf P}_\mu$ and ${\mathbf M}_{\mu\nu}$ are the generators of the extended DFR algebra.  These relations are important because they are essential for the construction of the Dirac equation in this extended DFR configuration $D=10$ $(x,\theta)$ space \cite{Amorim3}.  It can be shown that  the Clifford algebra structure generated by the 10 generalized Dirac matrices $\Gamma$ relies on these relations \cite{sigma}. In \cite{Amorim2}, the reader can find a complete discussion about the symmetries involved in this extended DFR space.

\section{Variational principles in curved spaces}\label{appendix}
\renewcommand{\theequation}{3.\arabic{equation}}
\setcounter{equation}{0}

In this section we will carry out a brief  review of the  formalism used  to calculate the equations of motion, i.e., the Euler-Lagrange (E-L) equations, when curved coordinates are introduced,  i.e., the vielbein method. 

We will not discuss here the features of physical definition of particle in curved space \cite{bd,wald,gibbons}.  Here, a particle  is an object which configuration space is a pseudo Riemannian manifold $({\cal M},g)$, the spacetime manifold.  The particle dynamics is defined by constraints and Hamiltonian.

The fundamental structure necessary on ${\cal M}$ in order to formulate Hamiltonian mechanics is a symplectic form $\omega_{\mu\nu}$, which is non-degenerate (it has a unique inverse, $\omega^{\mu\nu}$ satisfying $\omega^{\mu\nu}\,\omega_{\mu\nu}\,=\,\delta^\mu_\rho$), closed 2-form on ${\cal M}$, i.e., $\omega$ is a tensor field of type (0,2) on ${\cal M}$.  For any tangent vector $v^{\nu}$ on ${\cal M}$, we have $\omega_{\mu\nu}\,v^{\mu}\,=\,0$, if and only if $v^{\mu}=0$.  We will talk more about this symplectic form here in the future.

Let us consider a system represented by a Lagrangian function $L$ which depends on two sets of independent  local variables, the local coordinates $\{x^a\} $ and the extra coordinates $z=\{\chi^a , \xi_a \}$ (these variables are also called Einstein-Kramer variables \cite{ek,Amorim}), where $a,b,\dots$ denote Lorentz-indices running from $0, \dots, D$, with the associated  action
 \begin{equation}\label{action}
S_{local}=\int^{\tau_2}_{\tau1} d\tau L\left( x^a, \dot x^a, z, \dot z\right) .
 \end{equation}
Notice that we are in Minkowski spacetime for the time being.
Using the principle of least action with Dirichlet boundary conditions in both sets of variables, we can write that,
$$\delta x^a(\tau_1)=\delta x^a(\tau_2)=\delta z(\tau_1)=\delta z(\tau_2)=0\,\,,$$ 
and the E-L equations are established as
\begin{eqnarray}\label{e-l}
\frac{\partial L}{\partial x^a} &-& \frac{d}{d\tau} \frac{\partial L}{\partial \dot x^a} =0, \nonumber \\
\frac{\partial L}{\partial z} &-& \frac{d}{d\tau} \frac{\partial L}{\partial \dot z} =0\,\,.
\end{eqnarray}
On the other hand,  the independence between the variables can be written through the relations
\begin{equation}\label{eqn1}
\frac{\partial z}{\partial x^a} =0, \quad \frac{\partial \dot z}{\partial x^a} =0, \quad \frac{\partial \dot x^a}{\partial x^a}=0\,\,.
\end{equation}
The relation between the variables defined in the local frame with Lorentz-indices and the variables defined in the coordinate basis, with world-indices $\mu,\nu,\dots$ running from $0, \dots, D$, is given by the intertwining of the local coordinate axes named vielbein (vierbeins or tetrads) matrices at each point X in spacetime and then project all tensor quantities onto these local, Lorentzian inertial frame axes.  The vielbeins $V^a_{\ \ \mu}$ and $V^\nu_{\ \ a}$ are defined by,
\begin{eqnarray}\label{vielbein}
&&V^\mu_{\ \ a} V^b_{\ \ \mu} = \delta^b_a, \qquad \qquad \qquad V^a_{\ \ \mu} V^\nu_{\ \ a} = \delta^{\nu}_{\mu}, \nonumber \\
&&\eta_{ab}V^a_{\ \ \mu} V^b_{\ \ \nu}=g_{\mu\nu}\qquad \qquad \qquad \eta^{ab}V_a^{\ \ \mu} V_b^{\ \ \nu}=g^{\mu\nu}\,\,,
\end{eqnarray}
where $\eta_{ab}$ is the Minkowski metric $(+---)$.  So, the flat and curved variables are related to each other as
\begin{eqnarray}\label{vel1}
&&\dot x^a =V^a_{\ \ \mu} \dot x^\mu, \qquad \qquad \qquad \chi^a= V^a_{\ \ \mu} \chi^\mu \nonumber \\
&&\qquad \qquad \qquad\xi_a=V^\mu_{\ \ a} \xi_\mu\,\,.
\end{eqnarray} 
Here the variables $\chi^a$ and $\dot x^a$ can be seen as components of vector fields and $\xi_a$ are the components of a co-vector field, i.e., $\xi_a \in T^*{\cal M}$. 
%
The transformation rules between the flat and curved indices for $\dot z$  involve the following covariant generalizations,
\begin{eqnarray}\label{vel2}
\dot \chi^a=V^a_{\  \  \mu} \frac{D}{D\tau} \chi^\mu=V^a_{\  \  \mu} \dot x^\beta \nabla_\beta \chi^\mu \nonumber \\
\dot \xi_a= V^\mu_{\  \ a}\frac{D}{D\tau} \xi_\mu=V^\mu_{\  \ a}\dot x^\beta\nabla_\beta \xi_\mu,
\end{eqnarray}
where we used the relations in (\ref{eqn1}) and the definition,
\begin{equation}\label{eqn2}
\frac{D}{D\tau} A^\mu_{\  \ \nu}=\dot A^\mu_{\  \ \nu} + \Gamma^\mu_{\alpha\beta}A^\alpha_{\  \ \nu}\dot x^\beta - \Gamma^\alpha_{\nu \beta}A^\mu_{\  \ \alpha}\dot x^\beta\,=\,\dot{x}^\beta\nabla_\beta\,A^\mu_{\ \nu} 
\equiv \;\stackrel{\circ}{A^\mu}_{\ \nu}\,\,,
\end{equation}
where $\nabla_\mu$ is the usual covariant derivative.  We also suppose that the connection is symmetric, i.e., $\Gamma^{\sigma}_{\mu\nu}\,=\,\Gamma^\sigma_{\nu\mu}$, which implies that the we are in a torsion free regime. 

With these relations we can show that the conditions  (\ref{eqn1}) in the coordinate basis become
\begin{eqnarray} \label{con2}
\nabla_\mu\chi^\nu=0,&& \quad \nabla_\mu \xi_\nu = 0\,\,, \nonumber \\
\nabla_\mu\dot\chi^\nu=0,&& \quad \nabla_\mu \dot\xi_\nu =0\,\,.
\end{eqnarray}
We need to be careful because even if the above relations are zero their covariant derivatives (second order covariant derivatives) are not zero necessarily. An important and useful  relation between the vielbein matrices and the Christoffel symbols are
\begin{equation}\label{eqn4}
\frac{\partial V^a_{\  \  \mu}}{\partial x^\nu}=V^a_{\ \ \lambda} \Gamma^\lambda_{\mu\nu}, \qquad \qquad \qquad \frac{\partial V^\mu_{\ \ a}}{\partial x^\nu}=-V^\lambda_{\ \ a} \Gamma^\mu_{\lambda\nu}\,\,.
\end{equation}
With the rules (\ref{vielbein}), (\ref{vel1}) and (\ref{vel2}), the Lorentz invariant quantities in the local basis become quantities which are invariant under the general transformation of coordinates in the coordinate  basis. The functional (\ref{action}) will rely on variables with holonomic indices
\begin{equation}\label{action1}
S=\int^{\tau_2}_{\tau1} d\tau L\left( x^\mu, \dot x^\mu, z^A,  \stackrel{\circ}{z}^A\right) 
\end{equation}
and we are using  the short notations  $z^A=(\chi^\mu,\xi_\mu)$   and   $\stackrel {\circ}{z}\equiv Dz/D\tau$ that have been defined in (\ref{eqn2}). In order to calculate the E-L equations for this general covariant action  we will use the variational principle. Let us start  with the  variation of the functional (\ref{action1})  
\begin{equation}\label{var3}
\delta S= \int^{\tau_2}_{\tau1} d\tau \left[ \frac{\partial L}{\partial x^\mu} \delta x^\mu + \frac{\partial L}{\partial \dot x^\mu} \delta \dot x^\mu  +  \frac{\partial L}{\partial z^A} \delta z^A + \frac{\partial L}{\partial \stackrel{\circ}{z}^A} \delta \stackrel{\circ}{z}^A \right]\,\,.
\end{equation} 
In the usual formalism, to derive the E-L equations (\ref{e-l}), besides the vanishing of $\delta S$ and the Dirichlet conditions,  the variation and the ``time'' derivative commute  
$$\delta \frac{d}{d\tau} x^a=\frac{d}{d\tau} \delta x^a \qquad \mbox{and} \qquad  \delta \frac{d}{d\tau} z=\frac{d}{d\tau}\delta z\,\,.$$ 
%
 %
In our case  we perform the variation $\delta \dot x^\mu$  using  the local basis and the vielbein matrices.  From (\ref{vel1}) the variation $\delta x^a$ can be written as follows
\begin{equation}
\delta \dot x^a =\frac{d}{d\tau} \delta x^a=\frac{d}{d\tau} \left( V^a_{\ \ \mu} \delta x^\mu \right)\,\,.
\end{equation}
Using the relations (\ref{eqn4}) and  the fact that $\delta \dot x^\mu=V^\mu_{\ \ a} \delta \dot x^a$ the variation of $\dot x^\mu$ is equal to
 \begin{equation}\label{var1}
 \delta \dot x^\mu = \frac{d}{d\tau} \delta x^\mu + \Gamma^\mu_{\alpha\beta} \delta x^\alpha \dot x^\beta= \frac{D}{D\tau} \delta x^\mu\,\,,
 \end{equation}
where we used that $\Gamma^\sigma_{\mu\nu}\,=\,V^\sigma_\lambda\,\Gamma^\lambda_{\mu\nu}$.
 Proceeding in a similar way for  $\delta \dot z^A$  it is easy to show that the variations $\delta\stackrel{\circ}{\chi^\mu}$ and $\delta\stackrel{\circ}{\xi}_\mu $ have the following form
 \begin{eqnarray}\label{var2}
 \delta \stackrel{\circ}{\chi^\mu} &=& \frac{d}{d\tau} \delta \chi^\mu + \Gamma^\mu_{\alpha\beta} \delta \chi^\alpha \dot x^\beta \nonumber \\
  \delta \stackrel{\circ}{\xi}_\mu &=& \frac{d}{d\tau} \delta \xi_\mu - \Gamma^\alpha_{\mu\beta} \delta \xi_\alpha \dot x^\beta.
 \end{eqnarray} 
 Substituting relations (\ref{var1}) and (\ref{var2}) into the integral in (\ref{var3}) and integrating by parts using the Dirichlet conditions ($\delta x^\mu(\tau_1)=\delta x^\mu(\tau_2)=\delta z^A(\tau_1)=\delta z^A(\tau_2)=0 $) 
we obtain the covariant  E-L equations of motion for $\{ x^\mu \}$ 
 \begin{equation}\label{E-L4}
 \nabla_\mu L - \frac{D}{D\tau} \left( \frac{\partial L}{\partial \dot x^\mu} \right) =
 \nabla_\mu L - \dot x^\sigma\nabla_\sigma \left( \frac{\partial L}{\partial \dot x^\mu} \right) =0
 \end{equation}
 and for $z^A$ we have that
 \begin{eqnarray}\label{E-L3}
 \frac{\partial L}{\partial \chi^\mu} -\frac{D}{D\tau} \left(\frac{\partial L}{\partial \stackrel{\circ}{\chi}^\mu} \right) 
&=&\frac{\partial L}{\partial \chi^\mu} -\dot x^\sigma\nabla_\sigma \left(\frac{\partial L}{\partial \stackrel{\circ}{\chi}^\mu} \right)=0  
\end{eqnarray}
and
\begin{eqnarray}
\frac{\partial L}{\partial \xi_\mu} -\frac{D}{D\tau} \left( \frac{\partial L}{\partial \stackrel{\circ}{\xi}_\mu} \right) &=&\frac{\partial L}{\partial \xi_\mu} -\dot x^\sigma\nabla_\sigma\left( \frac{\partial L}{\partial \stackrel{\circ}{\xi}_\mu} \right)=0\,\,. 
 \end{eqnarray}
The star product, mentioned in the first section, for example, can be covariantly generalized as
\begin{eqnarray}
f\star g &=& f \exp\left( \overleftarrow{\nabla}_\mu
\frac i2 \omega^{\mu\nu} \overrightarrow{\nabla}_\nu
\right)g \nonumber\\
&=&\sum_{n=0}^{\infty} \frac 1{n!}\left( \frac i2\right)^n
\omega^{\mu_1\nu_1}\dots \omega^{\mu_n\nu_n}
(\nabla_{\mu_1}\dots \nabla_{\mu_n} f)
(\nabla_{\nu_1}\dots \nabla_{\nu_n} g)\,.
\label{star}
\end{eqnarray}
where $\omega$ is a closed non-degenerate 2-form \cite{vass}.  Geometrically speaking, in few words, let 
$\pi:\,T^*{\cal M} \rightarrow {\cal M}$ be the canonical projection, where $T^*{\cal M}$ is the cotangent bundle over the spacetime manifold ${\cal M}$ with the symplectic 2-form $\omega$.  In a local chart ($\pi^{-1}(U),x^\mu,p_\mu$), where $U\subseteq{\cal M}$ and the 2-form $\omega$ is given by the formula \cite{kalinin}

$$\omega\,=\,\sum_\mu\,dp_\mu \wedge dx^\mu\,\,.$$

\ni Therefore, we can say that ${\cal M}$ is equipped with a closed non-degenerate 2-form $\omega$ which, in a local coordinate system,

$$\partial_\mu\,\omega_{\nu\rho}\,+\,\partial_\rho\,\omega_{\mu\nu}\,+\,\partial_\nu\,\omega_{\rho\mu}\,=\,0\,\,,$$

\ni and this implies that $\omega$ satisfies the Jacobi identities.

Now we can see clearly the geometrical meaning of EK variables $z=\{\chi^a,\xi_a\}$, where $\chi^a\,\in\,{\cal M},\,\xi_a\in\,T^*{\cal M}$ and $(\xi_1,\ldots,\xi_n)$ as said before, are the components of the cotangent vectors in the coordinate basis associated with $(\chi^1,\dots,\chi^n)$.  In terms of $(\chi^a,\xi_a)$ the symplectic 2-form $\omega$ (defined before) can be defined as

$$\omega\,=\,\sum_\mu\,d\xi_a\,\wedge\,d\chi^a\,\,,$$

\ni and $\omega$ can be used to construct the Poisson brackets of the system.

For high order tensors the covariant equations can also be generalized. As a simple example let us consider the first order action,
\begin{equation}\label{action4}
S=\int d\tau\left[ p_\mu \dot x^\mu -\frac{1}{2}(g^{\mu\nu}p_\mu p_\nu +m^2)\right]
\end{equation}
that is just the relativistic first order action for a  particle in curved space. From the E-L equations (\ref{E-L4}) it is easy to verify that
\begin{equation}\label{eqn6}
\dot x^\mu \nabla_\mu p_\nu =0\,\,,
\end{equation}
meaning momentum conservation $\frac{Dp_\nu}{D\tau}=0$. And, for  $p_\mu$ the equations of motion lead us  to the relation $p_\mu=g_{\mu\nu}\dot x^\nu$, which is just the momentum definition in this case.  Inserting it in (\ref{eqn6}) the geodesic equations can be obtained as
\begin{equation}
\frac{D \dot x^\mu}{D\tau}=\ddot x^\mu+\Gamma^\mu_{\alpha\beta} \dot x^\alpha \dot x^\beta =0
\end{equation}
as expected. The covariant E-L equations can be computed directly by using the non covariant ones and the vielbein matrices,  using the chain rule in order to transform the derivatives relative to the nonholonomic variables into the holonomic ones. The variational principle was introduced in order to clarify the validity of this equations.

Concerning the symmetries of (\ref{action4}), let us consider an infinitesimal  transformation of coordinates $\delta x^\mu=x'^\mu - x^\mu =\epsilon^\mu(x)$ which implies an infinitesimal metric tensor transformation,   
\begin{equation} \label{diff1}
\delta g_{\mu\nu}= -\nabla_\mu \epsilon_\nu -\nabla_\nu \epsilon_\mu\,\,.
\end{equation}
The action (\ref{action4}) is invariant under diffeomorphism  if $\dot x^\mu$ and the momenta $p_\mu$ transform like
\begin{eqnarray}\label{diff3}
\delta \dot x^\mu &=& \frac{D}{D\tau} \delta x^\mu =\dot x^\alpha \nabla_\alpha \epsilon^\mu\,\,, \nonumber \\
\delta p_\mu &=&- p_\alpha\nabla_\mu \epsilon^\alpha\,\,,
\end{eqnarray}
equivalently the rule transformation of the momenta can be written as $\delta p^\mu =p^\alpha\nabla_\alpha \epsilon^\mu$.
\section{Noncommutative Free Particle}
\renewcommand{\theequation}{4.\arabic{equation}}
\setcounter{equation}{0}

In this section,  we present a  curved space generalization of the algebraic structure  found in \cite{Amorim4} with relativistic version in \cite{Amorim6}. To achieve this goal, it is introduced a  constrained Hamiltonian system  living in a phase space spanned by the quantities $x^\mu,Z^\mu$ and $\theta^{\mu\nu}$ and their conjugate momenta $p_\mu,K_\mu$ and $\pi_{\mu\nu}$, respectively.  The coordinates $x^\mu$ represent the curved coordinates, $\theta^{\mu\nu}$, as has been said, is the object of noncommutativity which is considered as an independent coordinate  in the phase space and $Z^\mu$
represents auxiliary variables introduced in order to properly implement  space-time noncommutativity.

In \cite{Amorim5} the first order action which generates all the noncommutative algebra preserving the Lorentz symmetry is found, the generalization to curved spaces is almost direct and is given by
\begin{equation}
\label{26}
S=\int d\tau L_{FO}=\int d\tau\left[ p\cdot\dot x+K\cdot\stackrel{\circ}{ Z}+\pi\cdot\stackrel{\circ}{\theta}-\lambda_a\,\Xi^a-\lambda\,\Upsilon \right]
\end{equation}
the center dots mean contraction of the indices (internal products) using the metric tensor $g_{\mu\nu}(x)$ which depends only on the curved coordinates, the symbol $^\circ$  means time-covariant derivative (\ref{eqn2}) which is necessary in order to preserve the invariance under general transformation of coordinates. The $2(D+1)$ second class constraints $\Xi^a=(\Psi^\mu,\Phi_\mu)$ \cite{Amorim5} are given by
\begin{eqnarray}
\label{11}
\Psi^\mu&=&Z^\mu-{1\over2}\theta^{\mu\nu}p_\nu
\nonumber\\
\Phi_\mu&=&K_\mu-p_\mu,
\end{eqnarray}
where $\lambda_a=(\lambda_{1\mu}, \lambda_2^{\mu})$ in (\ref{26}) are the Lagrange multipliers. 

By starting from the first order Lagrangian (\ref{26}),we arrive at 
\begin{equation}
\label{29a}
L_1=p\cdot(\dot x+\dot Z)+\pi\cdot \dot \theta-\lambda_1\cdot (Z-{1\over2}\theta\cdot p)\,-\,\lambda_2\cdot \,(\,p\,-\,K\,)
\,-\,\gamma({1\over{\lambda^2}}\pi^2+p^2+m^2)
\end{equation}

\noindent if one uses the equation of motion for $\lambda_2^\mu$, which is $p^\mu-K^\mu=0$. We observe  that the form of the first class constraint is also simplified due to symmetry. If we now use the equation of motion for $\lambda_{1\,\mu}$, which is just $Z^\mu - {1\over2}\theta^{\mu\nu}p_\mu=0$, we arrive at

\begin{equation}
\label{29b}
L_2=p\cdot \dot x+\pi\cdot \dot \theta-\gamma({1\over{\lambda^2}}\pi^2+p^2+m^2)+{1\over2}p\cdot \theta\cdot \dot p
\end{equation}

In \cite{Amorim6} the first class constraint $\Upsilon$  generated appropriate transformations for all variables (reparametrization invariance) of the extended phase space and has  the form 
\begin{equation}
\label{25}
\Upsilon={1\over{\kappa^2}}\chi'+\chi,
\end{equation}
in our treatment the first class quantities $\chi$ and $\chi'$ in a curved space are expressed as 
\begin{equation}
\label{3}
\chi={1\over2}(g^{\mu\nu} p_\mu p_\nu+m^2)
\end{equation}
and
\begin{equation}
\label{23}
\chi'\,=\, \frac 12 \Big\{\pi\cdot \pi\,+\,\pi\cdot K\cdot p\,+\,\frac 14 \Big[(K\cdot K)(p\cdot p)\,-\,(K\cdot p)^2\Big]\Big\}
\end{equation}
The  commutation relations under the Poisson brackets for the phase space variables are the usual ones
\begin{eqnarray}
\label{10}
\{x^\mu,p_\nu\}&=&\delta^\mu_\nu
\nonumber\\
\{\theta^{\mu\nu},\pi_{\rho\sigma}\}&=&\delta^{\mu\nu}_{\,\,\,\,\rho\sigma}\nonumber\\
\{Z^\mu,K_\nu\}&=&\delta^\mu_\nu
\end{eqnarray}
where $\delta^{\mu\nu}_{\,\,\,\,\rho\sigma}=\delta^\mu_\rho\delta^\nu_\sigma-\delta^\mu_\sigma\delta^\nu_\rho$.  Using the standard Dirac's procedure for treatment of singular systems \cite{Dirac} it is easy to verify that action (\ref{26}) has a first class constraint $\Upsilon$ since

\begin{equation}
\{ \Upsilon , \Psi^\mu \}=0, \quad \{ \Upsilon , \Phi_\mu \}=0
\end{equation}
and the second class set of constraints $\Xi^a$ with the associated non degenerated second class constraint matrix
\begin{equation}
\label{12}
(\Delta^{ab})=
\pmatrix{\{\Psi^\mu,\Psi^\nu\}&\{\Psi^\mu,\Phi^\nu\}\cr
	\{\Phi^\mu,\Psi^\nu\}&\{\Phi^\mu,\Phi^\nu\}}
=
\pmatrix{0&g^{\mu\nu}\cr
	-g^{\mu\nu}&0} 
\end{equation}
with  inverse 
\begin{equation}
\label{13}
(\Delta^{-1}_{ab})=
\pmatrix{0&-g_{\mu\nu}\cr
	g_{\mu\nu}&0}
\end{equation}
As well known the Dirac brackets between any two phase space functions $A$ and $B$ is given by \cite{Dirac}
\begin{equation}
\label{14}
\{A,B\}_D=\{A,B\}-\{A,\Xi^a\}\Delta^{-1}_{ab}\{\Xi^b,B\}
\end{equation}
As one can verify, the algebraic structure above permits to compute the Dirac brackets
\begin{equation}
\label{15}
\matrix{\{x^\mu,x^\nu\}_D=\theta^{\mu\nu}&\,\,\,\,\,\{x^\mu,p_\nu\}_D=\delta^\mu_\nu\cr
        \{p_\mu,p_\nu\}_D=0& \,\,\,\,\,\{\theta^{\mu\nu},\pi_{\rho\sigma}\}_D=\delta^{\mu\nu}_{\,\,\,\rho\sigma}\cr
        \{\theta^{\mu\nu},\theta^{\rho\sigma}\}_D=0&\,\,\,\,\,\{\pi_{\mu\nu},\pi_{\rho\sigma}\}_D=0\cr
        \{x^\mu,\theta^{\rho\sigma}\}_D=0&\,\,\,\,\,\,\,\,  \{x^\mu,\pi_{\rho\sigma}\}_D=-{1\over2}\delta^{\mu\nu}_{\,\,\,\rho\sigma}p_\nu\cr
        \{p_\mu,\theta^{\rho\sigma}\}_D=0&\,\,\,\,\,\{p_\mu,\pi_{\rho\sigma}\}_D=0}
\end{equation}
which involves the physical variables $x^\mu, p_\mu, \theta^{\mu\nu}$ and $\pi_{\mu\nu}$. The first relation show the noncommutative character of the model. The brackets listed above   generalize the algebra found in Ref.\cite{Amorim1,Amorim2}. It is also interesting to display the remaining Dirac brackets where 
the auxiliary variables $Z^\mu$ and $K_\mu$ appear:

\begin{equation}
\label{16}
\matrix{\{Z^\mu,K_\nu\}_D=0&\,\,\,\,\,\{Z^\mu,Z^\nu\}_D=0\cr
        \{K_\mu,K_\nu\}_D=0&\,\,\,\,\,\{Z^\mu,x^\nu\}_D=-{1\over2}\theta^{\mu\nu}\cr
        \{K_\mu,x^\nu\}_D=-\delta_\mu^\nu&\,\,\,\,\,\{Z^\mu,p_\nu\}_D=0\cr
        \{K_\mu,p_\nu\}_D=0&\,\,\,\,\,\{Z^\mu,\theta^{\sigma\rho}\}_D=0\cr
        \{Z^\mu,\pi_{\sigma\rho}\}_D={1\over2}\delta^{\mu\nu}_{\,\,\,\sigma\rho}p_\nu&\,\,\,\,\,\{K^\mu,\theta^{\sigma\rho}\}_D=0\cr
        \{K_\mu,\pi_{\sigma\rho}\}_D=0&{}}
\end{equation}
An important quantity  is the shifted coordinate 
\begin{equation}
\label{17}
X^\mu=x^\mu+{1\over2}\theta^{\mu\nu}p_\nu
\end{equation}
as can be verified such coordinates fulfill the canonical relations
\begin{equation}
\label{18}
\matrix{\{X^\mu,X^\nu\}_D=0&\,\,\,\,\,\{X^\mu,p_\nu\}_D=\delta^\mu_\nu\cr
        \{X^\mu,x^\nu\}_D={1\over2} \theta^{\mu\nu}& \,\,\,\,\,\{X^\mu\theta^{\rho\sigma},\pi_{\rho\sigma}\}_D=0\cr
        \{X^\mu,\pi_{\rho\sigma}\}_D=0&\,\,\,\,\,\{X^\mu,Z^\nu\}_D=-{1\over2} \theta^{\mu\nu}\cr
        \{X^\mu,K_\nu\}_D=\delta^\mu_\nu&\,\,\,\,\,\,\,\,  }0
\end{equation}
this quantity play a fundamental role in the relativistic treatment where the generator of the Poincar\'e group $M^{\mu\nu}$ is defined using an appropriated combinations of the shifted coordinate, momenta and NC coordinates and their conjugate momenta \cite{Amorim6}.

As in the ordinary case, it is also possible here to eliminate some of the variables in 
 favor of the others, by using for instance the second class constraints in a strong way. By starting from the first order action (\ref{26}), we arrive at
 \begin{equation}
\label{29b}
S=\int d\tau\left[p\cdot\dot x+\pi\cdot\stackrel{\circ}{ \theta}+{1\over2}p\cdot\theta\cdot\stackrel{\circ}{p}-{\gamma\over2}\Upsilon \right]
\end{equation}
if one uses the equations of motion for $\lambda_{1\mu}$ and $\lambda_2^\mu$   which are $Z^\mu-\frac{1}{2}\theta^{\mu\nu}p_\nu$ and $p^\mu-K^\mu=0$ respectively. We observe  that the form of the first class constraint  in (\ref{29b}) is also simplified due to symmetry, then in the second class constraints surface $\Upsilon$ reduces to the simpler form
\begin{equation}
\label{31}
\Upsilon={1\over2}\left(g^{\mu\nu}p_\mu p_\nu+{1\over{\kappa^2}}g^{\mu\alpha}g^{\nu\beta}\pi_{\mu\nu}\pi_{\alpha\beta}+m^2\right).
\end{equation}
We observe that the last term in (\ref{29b}) has already appeared in  \cite{Deriglazov}. In the Deriglazov's treatment there is the introduction of a factor of $\theta^{-2}$ in the corresponding term in order to introduce an additional gauge invariance which can be fixed by imposing constant $\theta$'s. In those works there is no  term in $\pi$ and any dynamics for the $\theta$ sector, which is a necessary ingredient to implement the quoted symmetry. Also that symmetry  is broken if any interaction is introduced via minimal coupling procedures.

As can be verified, (\ref{29b}) can be the starting point for essentially the same structure described in
the first part of this  section. It is important to remark that  here the tensor fields $\theta^{\mu\nu}$ have been included as the objects of noncommutativity.   As a result  the counting of bosonic degrees of freedom are $D+1+\frac{D(D+1)}{2}$. This implies that in $D+1=4$ ($\{x^0,\dots,x^3\}$), the number of bosonic degrees of freedom would be $10$.

The invariance under diffeomorphism  is expected in gravity models.  In our model this symmetry can be retained. If the  transformations (\ref{diff1}) and (\ref{diff3}) hold, we can find  the   transformations of the remainder variables. For the NC objects and their momenta conjugate we found that,
\begin{eqnarray} \label{diff5}
\delta\theta^{\mu\nu}& =& \theta^{\mu\sigma}\nabla_{\sigma}\epsilon^\nu + \theta^{\sigma\nu}\nabla_\sigma\epsilon^\mu \nonumber \\
\delta \pi_{\mu\nu}&=& -\pi_{\sigma\nu}\nabla_\mu\epsilon^\sigma - \pi_{\mu\sigma}\nabla_\nu\epsilon^\sigma
\end{eqnarray}
and for the time-covariant derivatives
\begin{eqnarray}\label{diff7}
\delta \stackrel{\circ}{p}_\mu &=&- \stackrel{\circ}{p}_\sigma\nabla_\mu \epsilon^\sigma \nonumber\\
\delta \stackrel{\circ}{\theta}^{\mu\nu}&=& \stackrel{\circ}{\theta}^{\mu\sigma}\nabla_\sigma \epsilon^{\nu}+\stackrel{\circ}{\theta}\ ^{\sigma\nu}\nabla_\sigma \epsilon^{\mu}
\end{eqnarray}
and the Lagrange multiplier transforming as $\delta \gamma=0$. Using  (\ref{diff5}) and (\ref{diff7}) it is straightforward to show that  (\ref{29b}) is invariant under diffeomorphisms.

\section{Motion on Noncommutative phase space}
\renewcommand{\theequation}{5.\arabic{equation}}
\setcounter{equation}{0}

In the previous section it has been introduced the first order action used  to obtain the  constraint structure necessary to generate the Dirac brackets in the extended DFR phase space.   This structure is NC in the coordinates.
Now we will turn our attention to the dynamics obtained from (\ref{29b}). 

The E-L equations have to be calculated in a  covariant way (in section \ref{appendix} the treatment of variational principles in curved spaces was introduced).  In order to calculate the equations of motion related to the coordinates $x^\mu$ it is useful to rewrite the Lagrangian as
\begin{equation}\label{30b}
L_{FO}=p\cdot\dot x+\pi_{\mu\nu} \dot x^\alpha\nabla_\alpha\theta^{\mu\nu}+{1\over2}p_\mu\theta^{\mu\nu} \dot x^\alpha \nabla_\alpha p_\nu-{\gamma\over2}\Upsilon
\end{equation}
where we write $L_{FO}$ in terms of the covariant derivative $\nabla_\mu$ and (\ref{eqn2}) was used. From (\ref{E-L4}) the  first set of equations, related with the coordinates $\{x^\mu\}$, is given by
\begin{equation}
\frac{Dp_\mu}{D\tau}  +{1\over 2}p_\alpha\theta^{\alpha\beta}\dot x^\sigma [\nabla_\sigma , \nabla_\mu]  p_\beta + \pi_{\alpha\beta} \dot x^\sigma[\nabla_\sigma , \nabla_\mu ] \theta^{\alpha\beta} = 0\,\,,
\end{equation}
the commutator of the covariant derivatives can be related with the curvature tensor $R^\lambda_{ \  \ \mu\nu\kappa}$ ($[\nabla_\mu , \nabla_\nu]A_\sigma = - R^\lambda_{\  \ \sigma\mu\nu} A_\lambda$ and similar relations for high order tensors) left the last equation in the form
\begin{equation}\label{eqn-nc1}
{Dp_\mu\over D\tau}  -{1\over2}\left( p_\alpha \theta^{\alpha\beta}p_\sigma -4 \pi_{\alpha\sigma}\theta^{\alpha \beta}\right)R^{\sigma}_{\  \ \beta\lambda\mu}\dot x^\lambda =0\,\,.
\end{equation}
Here  it is important remember that  even $\nabla_\alpha p_\mu=0$ and $\nabla_\alpha \theta^{\mu\nu}=0$ are true, they are  relations and are not identities, then the second  covariant  derivatives $\nabla_\beta\nabla_\alpha  p_\mu$ and $\nabla_\beta\nabla_\alpha \theta^{\mu\nu}$ in general are different to zero.
For the momenta $p_\mu$ the E-L equations resulting  from (\ref{E-L3}) are
\begin{equation}\label{eqn-nc2}
\dot x^\mu + \theta^{\mu\alpha}\stackrel{\circ}{p}_\alpha +{1\over 2} \stackrel{\circ}{\theta}^{\mu\alpha}p_\alpha - \lambda g^{\mu\nu}p_\nu =0
\end{equation}
For the extended coordinates $\theta$ the E-L equations  resulting are,
\begin{equation}
{D\pi_{\mu\nu} \over D\tau} - {1\over 4} p_{[\mu} \stackrel{\circ}{p}_{\nu]}=0\,\,,
\end{equation}
where the square brackets means antisymmetry in the indices and for its  momenta conjugate we find
\begin{equation}\label{eqn-nc4}
\kappa^2{D\theta^{\mu\nu}\over D\tau} -\lambda g^{\mu\alpha}g^{\nu\beta}\pi_{\alpha\beta}=0\,\,.
\end{equation}
%
In the commutative limit, when $\theta^{\mu\nu} \rightarrow 0$, the equations (\ref{eqn-nc1}) and (\ref{eqn-nc2}) turn  into the corresponding equations for an ordinary particle in curved background introduced as an example at the end of  section \ref{appendix}. To implement the set of equations (\ref{eqn-nc1})-(\ref{eqn-nc4}) using some particular metric one could impose the corresponding symmetries to noncommutative variables, in this scheme the NC variables should be interpreted  as independent fields in the model. Another way to treat these equations is  using some dimensional reduction formalism  like Kaluza-Klein or holography.

In commutative spacetime, $g_{\mu\nu}$ is a classical ``c-number."  We can consider this approximation since quantum gravitational fluctuations can be ignored.  Besides, one does not take into account the back reaction to particle creation.  The curvatures necessary to create particles will necessarily build up during gravitational collapse or in the big bang.  These processes are not known in this extended NC spacetime.
As we have observed at the beginning of section 3, a detailed discussion about the particles which obey the equations of motion described above in curved extended NC $D=10$ spacetime is out of the scope of this work and it is subject for future analysis.
However, we can say that the trajectories described by these equations of motion live in this pseudo Riemannian manifold.  Besides we showed above that the particles dynamics can be defined from the construction of the NC Lagrangian constructed in ${\cal M}$.

\section{Conclusions}

There is some circumstances where the quantum effects of gravity itself has not a fundamental role.  In these circumstances, if we do not have an exact theory, it is believed that quantum field theory in curved spacetime should provide a good approximate description.

The effort to construct a theory that describes exactly the unification of general relativity and quantum mechanics tell us that exists a fundamental length.  At this point we can talk in terms of a NC spacetime, which has a fundamental length defined in Planck scale.


In this work a model of a noncommutative particle in curved space was constructed.  The action describing such particle was written into the extended Doplicher-Fredenhagen-Roberts spacetime, where now the noncommutative antisymmetric parameters $\theta^{\mu\nu}$ are incorporated as spacetime coordinates and therefore have conjugate momentum. We explored the results obtained in \cite{Amorim4} concerning the construction of an action that obeys the NC algebra and we have embedded this action in curved space.   It can be shown that this action can describe also the relativistic particles depicted in \cite{Amorim6,Deriglazov}.

\noindent Concerning the resulting algebra 
\begin{equation}
\{ x^\mu , x^\nu\}_D = \theta^{\mu\nu}(\tau)\,\,,
\end{equation}
the noncommutative coordinates are dependent on the ``time" since it is not a constant parameter as in the usual approaches.  Another way to incorporate non-constant noncommutative objects is using non-canonical symplectic two-forms in the context of symplectic mechanics \cite{gss}.

The resulting dimension of the spacetime  is $D+1+\frac{D(D+1)}{2}$ due to the new variables inclusion.  Another characteristic of our model is that, in the first order action describing the particle,  similar coupling terms between the noncommutative sector and the commutative one was reported by \cite{Deriglazov}, see equation (\ref{29b}) .

We review the main steps in curved space formalism and as said just above, embedded the NC action in this curved background.  To accomplish this we used the so-called Einstein-Kramer variables, which are auxiliary variables that divides the original vectorial space and the vielbein formalism which provides the relations between curved variables and the flat space ones.  The resulting action is invariant under diffeomorphism with the suitable rules of transformations of the NC variables. 

To describe the dynamics of this curved space NC system, we computed the Euler-Lagrange equations.  These equations have the correct commutative  limit at the surface where the noncommutative variables vanish $\theta^{\mu\nu}=0$. As a consequence of the incorporation of the new noncommutative variables, the curvature tensor appear in the covariant set of equations of motions. One deep analysis and applications  of the noncommutative motion  equations is beyond the scope of this work, research in this direction are being constructed and will be reported elsewhere.

As well known, linear wave equations in flat spacetime for arbitrary spin particles can be constructed.  An object for future investigation is to construct the basic structure of the NC curved spacetime propagation equations for spin 0 and 
$\frac 12$ particles and the subsidiary conditions.  

As another perspective, the minimal coupling or other generalizations to curved spacetime can introduce extra-subsidiary conditions or propagation equations.  However, these features are unknown in this extended NC background and can be investigated.

\end{document}